\documentclass[12pt]{article}
\usepackage[english]{babel}
\usepackage{epsfig}

\def\pin{$\mbox{}$\indent}  %force to indent the first paragraph

\def\bbt#1{\bibitem{#1} \label{bb:#1}}

\def\bsigma{\mbox{\boldmath $\sigma$}}
\def\bxi{\mbox{\boldmath $\xi$}}

\def\E{\mbox{\rm E}}
\def\Var{\mbox{\rm Var}}
\def\ustr#1#2{\;\,\stackrel{#1}{#2}\;\,}

\let\Journal=\it

\def\jpa#1{{\Journal J. Phys. A: Math. Gen.} {\bf #1}}
\def\jsp#1{{\Journal J. Stat. Phys.} {\bf #1}}

\def\pre#1{{\Journal Phys. Rev. E} {\bf #1}}

\begin{document}
\setcounter{page}{0}
\begin{titlepage}
\title{Local field dynamics in symmetric $Q$-Ising neural networks}
% author with addresses under the name
\author{ D.~Boll\'e
 	   \footnote{e-mail: desire.bolle@fys.kuleuven.ac.be.}
	   \footnote{Also at Interdisciplinair Centrum 
            voor Neurale Netwerken, K.U.Leuven, Belgium.} \\
	 Instituut voor Theoretische Fysica,
            K.U.\ Leuven, \\ B-3001 Leuven, Belgium \\ \\
         and G.~M.~Shim
           \footnote{e-mail: gmshim@cnu.ac.kr.}\\
	 Department of Physics, Chungnam National 
            University \\Yuseong, Taejon 305-764, R.O.~Korea}
\date{}
\maketitle
%\thispagestyle{empty}
%\newpage
\thispagestyle{empty}
\begin{abstract}
%\normalsize
\noindent
The time evolution of the local field in {\em symmetric} $Q$-Ising neural
networks is studied for arbitrary $Q$. In particular, the structure of
the noise and the appearance of gaps in the probability distribution are 
discussed. Results are presented for several values of $Q$ and compared 
with numerical simulations.
\end{abstract}
%PACS numbers: 87.10.+e, 02.50.+s, 64.60.Cn  \newline
\vspace*{2cm}
{\bf Key words:} Symmetric networks; $Q$-Ising neurons;
parallel dynamics; local field; probabilistic approach
%\vspace*{1cm}
%\noindent
\end{titlepage}

\section{Introduction}
\pin
In a number of papers in the nineties (cfr.~\cite{PZ}-\cite{BJS99} and  
references therein) the parallel dynamics of $Q$-Ising type
neural networks has been discussed for several architectures --extremely
diluted, layered feedforward, recurrent-- using a probabilistic
approach.  For the asymmetric extremely diluted and layered
architectures the dynamics can be solved exactly and it is known that
the local field only contains Gaussian noise. For networks with
symmetric  connections,
however, things are quite different. Even for extremely diluted versions
of these systems feedback correlations become essential 
from the second time step onwards, complicating the dynamics in a 
nontrivial way. 

A complete solution for the parallel dynamics of symmetric 
$Q$-Ising networks at zero-temperature taking
into account all feedback correlations, has been obtained only recently
using a probabilistic signal-to-noise ratio
analysis \cite{BJSF}-\cite{BJS99}.
Thereby it is seen that both for the fully connected  and  the extremely
diluted symmetric architectures, the local field contains a discrete and a
normally distributed noise part.  The difference between the two
architectures is that for the  diluted model
the discrete part at a certain time $t$ does not involve the
spins at all previous times $t-1, t-2, \ldots$ up to $0$ but only the 
spins at time step $t-1$.  Even so, this discrete part prevents
a closed-form solution of the dynamics but
a recursive scheme  can be developed in order to calculate the complete 
time evolution of
the order parameters, i.e.,  the retrieval overlap and the activity.

In the work above the focus has been on the non-equilibrium behavior of
the order parameters of the network. But, since the local field itself
is a basic ingredient in the development of the relevant recursive
scheme it is interesting to study also the non-equilibrium behavior
of the local field distribution. The more so since this distribution
does not 
convergence to a simple sum of Gaussians as is frequently thought, but it
develops a gap structure. This is precisely one of the points 
studied in detail in the present communication. Moreover, the analogies
and differences  between the fully connected architecture and the
symmetrically diluted one are highlighted. Finally, numerical
simulations are presented confirming the analytic study and giving
additional insight in the structure of these local field
distributions.

\section{The model}
\pin
Consider a neural network $\Lambda$ consisting of $N$ neurons which can take
values $\sigma_i$ {}from a discrete set
        $ {\cal S} = \lbrace -1 = s_1 < s_2 < \ldots < s_Q
                = +1 \rbrace $.
The $p$ patterns to be stored in this network are supposed to
be a collection of independent and identically distributed random
variables (i.i.d.r.v.), $\{{\xi}_i^\mu \in {\cal S}\}$,
$\mu \in {\cal P}=\{1,\ldots,p\}$ and   $i \in \Lambda$,
with zero mean, $E[\xi_i^\mu]=0$, and variance $A=\Var[\xi_i^\mu]$. The
latter is a measure for the activity of the patterns. 
Given the configuration
        ${\bsigma}_\Lambda(t)\equiv\{\sigma_j(t)\},
        j\in\Lambda=\{1,\ldots,N\}$,
the local field in neuron $i$ equals
\begin{equation}
        \label{eq:h}
        h_i({\bsigma}_{\Lambda}(t))=
                \sum_{j\in\Lambda} J_{ij}(t)\sigma_j(t)
\end{equation}
with $J_{ij}$ the synaptic coupling from neuron $j$ to neuron $i$.
In the sequel we write the shorthand notation $h_{\Lambda,i}(t) \equiv
h_i({\bsigma}_{\Lambda}(t))$.

For the extremely diluted symmetric (SED) and 
the fully connected (FC) architectures the  couplings are
given by the Hebb rule 
\begin{eqnarray}
     J_{ij}^{SED}&=&\frac{c_{ij}}{CA}
               \sum_{\mu \in {\cal P}} \xi_i^\mu \xi_j^\mu
     \quad \mbox{for} \quad i \not=j       \,, \quad J_{ii}^{SED}=0 \, ,
	\label{eq:JED}  \\
     J_{ij}^{FC}&=&\frac{1}{NA}
               \sum_{\mu \in {\cal P}} \xi_i^\mu \xi_j^\mu
     \quad \mbox{for} \quad i \not=j       \,, \quad J_{ii}^{FC}=0 \, ,
        \label{eq:JFC}  
\end{eqnarray}
with the $\{c_{ij}=0,1\}, i,j \in \Lambda$ chosen to be i.i.d.r.v. with
distribution
$\mbox{Pr}\{c_{ij}=x\}=(1-C/N)\delta_{x,0} + (C/N) \delta_{x,1}$ and
satisfying $c_{ij}=c_{ji} $.

For the diluted symmetric model the architecture is 
a local Cayley-tree  but, in contrast with the diluted asymmetric model,
it is no  longer directed such that it causes a feedback
from $t \geq 2$ onwards. In the limit
$N \rightarrow \infty$ the probability that the number of connections
$T_i=\{j\in \Lambda |c_{ij}=1\}$ giving information to the site
$i \in \Lambda$, is still a Poisson distribution with mean $C=E[|T_i|]$.
Thereby it is assumed that $ C \ll \log N$ and in order to get an infinite
average connectivity
allowing to store infinitely many patterns one also takes the limit $C
\rightarrow \infty$ \cite{BJS99}.

At zero temperature all neurons are updated in parallel according to the
rule
\begin{eqnarray}
        \label{eq:gain}
        \sigma_i(t+1) &  =   &
               \mbox{g}_b(h_{\Lambda,i}(t))
                  \nonumber      \\
               \mbox{g}_b(x) &\equiv& \sum_{k=1}^Qs_k
                        \left[\theta\left[b(s_{k+1}+s_k)-x\right]-
                              \theta\left[b(s_k+s_{k-1})-x\right]
                        \right]
\end{eqnarray}
with $s_0\equiv -\infty$ and $s_{Q+1}\equiv +\infty$. Here
$\mbox{g}_b(\cdot)$ is the gain function and $b>0$ is the gain parameter
of the system. For finite $Q$, this gain function is a step function.
The gain parameter $b$ controls the average slope of $\mbox{g}_b(\cdot)$.

\section{Local field dynamics}
\pin
In order to measure the retrieval quality of the system one can use the
Hamming
distance between a stored pattern and the microscopic state of the network
\begin{equation}
        d({\bxi}^\mu,{\bsigma}_\Lambda(t))\equiv
                \frac{1}{N}
                \sum_{i\in \Lambda}[\xi_i^\mu-\sigma_i(t)]^2         \,.
\end{equation}
This  introduces the main overlap and the arithmetic mean of the
neuron activities
\begin{equation}
        \label{eq:mdef}
        m_\Lambda^\mu(t)=\frac{1}{NA}
                \sum_{i\in\Lambda}\xi_i^\mu\sigma_i(t),
                \quad \mu \in {\cal P}\, ; \quad
        a_\Lambda(t)=\frac{1}{N}\sum_{i\in\Lambda}[\sigma_i(t)]^2    \,.
\end{equation}
The key question is then how these quantities evolve in time under the
parallel dynamics specified before.
For a general time step we find {}from eq.~(\ref{eq:gain}) using the law
of  large numbers (LLN) that in the thermodynamic limit  
\begin{eqnarray}
        m^1(t+1) \ustr{Pr}{=} \frac{1}{A} \langle\!\langle
                 \xi_i^1\mbox{g}_b(h_i(t)) \rangle\!\rangle , \quad
        a(t+1)   \ustr{Pr}{=} \langle\!\langle \mbox{g}_b^2(h_i(t))
                         \rangle\!\rangle \, ,
          \label{eq:a}
\end{eqnarray}
where the convergence is in probability \cite{SH}.
In the above $\langle\!\langle \cdot \rangle\!\rangle$
denotes the average both over the distribution of the embedded patterns
$\{\xi_i^\mu\}$ and the  initial configurations $\{\sigma_i(0)\}$. The
average over the latter is hidden in an average over the
local field through the updating rule (\ref{eq:gain}).

Some remarks are in order. For
the symmetric diluted model the sum over the sites $i$ is restricted to 
$T_j$, the part of the tree connected to neuron $j$. Moreover, for that
model the thermodynamic limit contains the limit 
$C \rightarrow \infty$
besides the $N \rightarrow \infty$ limit. In this
thermodynamic limit $C, N \rightarrow \infty$ all
averages have to be taken over the treelike structure, viz.
$\frac{1}{N}\sum_{i \in \Lambda} \rightarrow \frac{1}{C} \sum_{i \in
T_j}$, and the capacity defined by $\alpha =p/N$ has to be replaced by 
$\alpha =p/C$.

In (\ref{eq:a}) the local field is the main ingredient.
Suppose that the initial configuration of the network
$\{\sigma_i(0)\},{i\in\Lambda}$, is a collection of i.i.d.r.v.\ with mean
$\E[\sigma_i(0)]=0$, variance $\Var[\sigma_i(0)]=a_0$, and correlated with
only one stored pattern, say the first one $\{\xi^1_i\}$:
\begin{equation}
        \label{eq:init1}
        \E[\xi_i^\mu\sigma_j(0)]=\delta_{i,j}\delta_{\mu,1}m^1_0 A
\end{equation}
with $m^1_0>0$. By the LLN
one  gets for the main overlap and the activity at $t=0$
\begin{eqnarray}
        m^1(0)&\equiv&\lim_{(C),N \rightarrow \infty} m^1_\Lambda(0)
                \ustr{Pr}{=}\frac1A \E[\xi^1_i \sigma_i(0)]
                = m^1_0
                \label{eq:mo}       \\
        a(0)&\equiv&\lim_{(C),N \rightarrow \infty} a_\Lambda (0)
                \ustr{Pr}{=} \E[\sigma_i^2(0)]=a_0
                \label{eq:a0}
\end{eqnarray}
where the notation should be clear. In order to
obtain the configuration at $t=1$ we have to calculate the local
field (\ref{eq:h}) at $t=0$. To do this we employ the probabilistic  
signal-to-noise ratio analysis (\cite{PZ}-\cite{BJS99}). Recalling the   
learning rule (\ref{eq:JFC}) we separate the part containing 
the signal from the part containing the noise. 
In the  limit $N \rightarrow \infty$ we then arrive at
\begin{equation}
        h_i(0)
        \equiv
        \lim_{N \rightarrow \infty} h_{{\Lambda},i}(0)
        \stackrel{{\cal D}}{=}
        \xi_i^1 m^1(0) +  {\cal N}(0,\alpha a(0))
\label{eq:F16}
\end{equation}
where the convergence is in distribution \cite{SH} and
with ${\cal N}(0,V)$ representing a Gaussian random variable with
mean $0$ and variance $V$. We note that this structure of
the distribution of the local field at time zero -- signal plus Gaussian
noise -- is typical for all architectures treated in the literature.
 
For a general time step $t+1$, a tedious study reveals that the
distribution  of the local field is given by \cite{BJSF}, \cite{BJS99}
\begin{equation}
        h_i(t+1)=\xi_i^1m^1(t+1) + {\cal N}(0,\alpha a(t+1))
           + \chi(t) [F(h_i(t)-\xi_i^1m^1(t))+\alpha\sigma_i(t)]
              \label{eq:hrec}
\end{equation}
where $F=1$ for the fully connected architecture and $F=0$ for the 
symmetrically diluted one. So, the local field at time 
$t$ consists out of a discrete part and a normally distributed part, viz.
\begin{equation}
        h_i(t)=M_i(t) + {\cal N}(0, V(t))
\end{equation}
where $M_i(t)$  and $V(t)$ satisfy the recursion relations
\begin{eqnarray}
        && M_i(t+1)=\chi(t) [F(M_i(t)-\xi_i^1m^1(t))+\alpha\sigma_i(t)]
                         + \xi_i^1m^1(t+1)
     \label{eq:Mrec} \\
        \label{eq:Drec}
       && V(t+1)= \alpha a(t+1)A+F\chi^2(t)V(t)+
                2 F \alpha A \chi(t) {Cov}[\tilde r^\mu(t),r^\mu(t)] \,.
 \end{eqnarray}
The quantity $\chi (t)$ reads
\begin{equation}
        \chi(t) = \sum_{k=1}^{Q-1} f_{ h_i^\mu (t)}(b(s_{k+1}+s_k))
                   (s_{k+1}-s_k) 
            \label{eq:chi}
\end{equation}
where $f_{ h_i^\mu (t)}$ is the probability density of 
 $  h_i^\mu (t) $ in the thermodynamic limit.    
Furthermore, $r^\mu(t)$ is defined as
\begin{equation}
         r^\mu(t) \equiv \lim_{N \rightarrow \infty}
           \frac1{A\sqrt{N}}\sum_{i\in \Lambda} \xi_i^\mu
                \sigma_i(t), \quad   
           \mu \in {\cal P}\setminus\{1\}  \, ,      
	        \label{eq:w}
\end{equation}
and $\tilde r^\mu(t)$ is given by a similar expression with $\sigma_i(t)$
replaced by $\mbox{g}_b(h_{\Lambda,i}(t) -
                \frac{1}{\sqrt{N}}\xi_i^\mu r_\Lambda^\mu(t) )$.
Finally, as can be read off {}from eq.~(\ref{eq:Mrec}) the quantity
$M_i(t)$  consists out of a signal term and a discrete noise
term, viz.
\begin{equation}
        M_i(t)=\xi _i^1 m^1(t) + \alpha \chi(t-1)\sigma _i(t-1)
        + F\sum_{t'=0}^{t-2} \alpha
         \left[\prod_{s=t'}^{t-1} \chi(s)\right] \, \sigma _i(t')  \,.
         \label{eq:MM}
\end{equation}
Since different architectures contain different correlations not all
terms in these final equations are present, as is apparent through $F$. 
We remark that for the asymmetric diluted and the layered feedforward
architecture $M_i(t)=\xi _i^1 m^1(t)$ so that in these cases the local
field  consists
out of a signal term plus Gaussian noise for {\it all} time steps
\cite{BSVZ},\cite{BSV}. 

For the architectures treated here we still have to determine the  
probability density $f_{h_i(t)}$  in eq.~(\ref{eq:chi}).
This can
be done by looking at the form of $M_i(t)$ given by eq.~(\ref{eq:MM}).
The evolution equation tells us that $\sigma _i(t')$ can be replaced by
$g_b(h_i(t'-1))$ such that the second  and third terms of $M_i(t)$ are
the sums of stepfunctions of correlated variables. These are also
correlated through the dynamics with the normally distributed
part of $h_i(t)$. Therefore, the local field can be considered as a
transformation of a set of correlated normally distributed variables
$x_s$, which we choose to normalize. Defining the
correlation matrix $W = \left(\rho(s,s')\equiv \E[x_s x_{s'}] \right)$
we  arrive at the following expression for $f_{h_i(t)}$
for the fully connected model
\begin{eqnarray}
     f_{h_i(t)}(y)&=&\int dx_t \,\prod_{s=0}^{t-2} dx_s  ~
         \delta \left(y - M_i(t)-\sqrt{V(t)}\,x_t\right)
             \nonumber\\
             &\times& \frac{1}{\sqrt{\mbox{det}(2\pi W)}}
            ~\mbox{exp}\left(-\frac{1}{2}{\bf x} W^{-1}
            {\bf x}^T\right)
            \label{eq:fhdisfc}
\end{eqnarray}
with ${\bf x}=\{x_s\}=(x_0,\ldots x_{t-2},x_t)$. For the symmetric diluted
case this expression simplifies to
\begin{eqnarray}
     f_{h_i(t)}(y)&=&\int\prod_{s=0}^{[t/2]} dx_{t-2s} ~
         \delta \left(y -\xi^1_i m^1(t)- \alpha \chi(t-1)\sigma_i(t-1)
              -\sqrt{\alpha a(t)}\,x_t\right) \nonumber\\
             &\times& \frac{1}{\sqrt{\mbox{det}(2\pi W)}}
            ~\mbox{exp}\left(-\frac{1}{2}{\bf x} W^{-1}
            {\bf x}^T \right)
            \label{eq:fhdisd}
\end{eqnarray}
with ${\bf x}=(\{x_s\})=(x_{t-2[t/2]},\ldots x_{t-2},x_t)$. The brackets
$[t/2]$ denote the integer part of $t/2$.

\section{Gap structure}
\pin
The equilibrium distribution of the local field can be obtained by
eliminating the time dependence in the evolution equations (\ref{eq:hrec})
\begin{equation}
        \label{eq:hfix}
    h_i=\xi_i^1m^1 + \eta{\cal N}(0,\alpha a) +\alpha \chi \eta \sigma_i
\end{equation} 
with $\eta= 1/(1-\chi)$ for the fully connected architecture and
$\eta=1$ for the extremely diluted one. The 
corresponding updating rule (\ref{eq:gain}) 
\begin{equation}
    \sigma_i = g_b(\tilde {h_i} + \alpha \chi \eta \sigma_i) \, , 
    \quad \tilde{h_i} =  \xi_i^1 m_i^1 +  \eta{\cal N}(0,\alpha a)
    \label{eq: res1}
\end{equation}
in general admits more than one solution. A Maxwell construction
(see, e.g., refs. \cite{BJSF},\cite{BJS99},\cite{SF}) can be
made leading to a unique solution 
\begin{equation}   
       \sigma_i = g_{\tilde{b}}(\tilde{h_i})\, ,
       \quad \tilde{b}=  (b - \frac{\alpha \eta \chi}{2})
     \label{eq: res2}  
\end{equation}
such that we have
\begin{equation}
   \sigma_i = s_k \quad \mbox{if} \quad
     \tilde{b}(s_k+s_{k-1})+\alpha \chi\eta s_k 
                <  h_i <
     \tilde{b}(s_k+s_{k+1}) +\alpha \chi \eta s_k \, .
     \label{eq: res3}
\end{equation}     
for ${\tilde b} > 0 $. This unique solution can be used to  obtain
fixed-point equations for the main overlap and activity (\ref{eq:a}).   
Those equations which we choose not to write down explicitly here (see
refs. \cite{BJSF},\cite{BJS99})
are equal to the equations derived from a thermodynamic replica-symmetric 
mean-field theory approach \cite{BRS},\cite{BCS}. We remark that for
analog  networks ($Q \to
\infty$) such a Maxwell construction is not necessary because
eq.~(\ref{eq: res1}) has only one solution.

Next, we calculate the probability density of the local field  by
plugging this result (\ref{eq: res1})-(\ref{eq: res3}) into
(\ref{eq:hfix}) to obtain, forgetting about the site index $i$ and the
pattern index $1$
\begin{eqnarray}
   f(h) &=& \sum_{k=1}^Q  \frac{1}{\eta\sqrt{2\pi \alpha a}}
      \exp\biggl(
          -\frac{(h-\xi m -\alpha\chi\eta s_k)^2}{2\alpha a \eta^2}
         \biggr) \nonumber \\
           &\times&
      \biggl( \theta[ \tilde{b}(s_k+s_{k+1})+\alpha \chi\eta s_k -h]
             -\theta[ \tilde{b}(s_k+s_{k-1})+\alpha \chi\eta s_k -h]
       \biggr)
       \label{eq:distri}
\end{eqnarray}
meaning that (Q-1) gaps occur respectively at $ \tilde{b}(s_k+s_{k-1})+ 
\alpha\chi\eta s_{k-1} 
<  h < \tilde{b}(s_k+s_{k+1}) +\alpha\chi\eta s_k $ with width $\Delta h=
2\alpha\chi\eta/(Q-1)$. For analog networks no gaps occur. 
When ${\tilde b} \leq 0$ the effective gain function (\ref{eq: res2})
becomes two-state Ising-like as in the Hopfield model such that case only
one gap occurs. 

For $Q=2$ this expression simplifies to
\begin{eqnarray} 
  f(h) &=& \frac{1}{\eta \sqrt{2\pi \alpha a}} \exp\biggl(
             -\frac{(h-\xi m -\alpha\chi\eta )^2}{2\alpha a \eta^2}
          \biggr) \theta(h-\alpha\chi\eta) \nonumber \\
       &+&
          \frac{1}{\eta\sqrt{2\pi \alpha a}} \exp\biggl(
             -\frac{(h-\xi m +\alpha\chi\eta )^2}{2\alpha a \eta^2}
          \biggr) \theta(-h-\alpha\chi\eta) 	  
\end{eqnarray}	  
and for Q=3 we have
\begin{eqnarray}
    f(h) &=& \frac{1}{\eta\sqrt{2\pi \alpha a}} \exp\biggl(
             -\frac{(h-\xi m -\alpha\chi\eta )^2}{2\alpha a \eta^2}
          \biggr) \theta(h-\tilde{b}-\alpha\chi\eta) \nonumber \\
       &+&
          \frac{1}{\eta\sqrt{2\pi \alpha a}} \exp\biggl(
             -\frac{(h-\xi m )^2}{2\alpha a \eta^2}
          \biggr) \theta(\tilde{b}^2-h^2) \nonumber \\
       &+&
          \frac{1}{\eta\sqrt{2\pi \alpha a}} \exp\biggl(
             -\frac{(h-\xi m +\alpha\chi\eta )^2}{2\alpha a \eta^2}
          \biggr) \theta(-\tilde{b}-\alpha\chi\eta-h) \, .
\end{eqnarray}
Similar formula can be written down for bigger values of $Q$. For $Q=2$
this result seems to be consistent with the gap in the internal-field
distribution for an infinite range spin glass found by
a Bethe-Peierls-Weiss approach \cite{SK} (see also~\cite{ZC}-\cite{CS}).

We have investigated this probability distribution  numerically 
using the corresponding fixed-point equations  mentioned before, for
several  values of $Q$ and compared them with those obtained from  
numerical simulations of the dynamics for networks of $N=6000$ neurons. 
Some typical results are shown in  figs.~1-6.

In figs.~1-2 the local field distribution for the fully connected $Q=2$ 
network is shown
for a retrieval state ($\alpha=0.13,m_0=0.5$) just below the critical   
capacity and a non-retrieval spin-glass state ($\alpha=0.14, m_0=0.2$)
just above it. Both the first few time steps
and the equilibrium result derived above are compared with  
numerical simulations. They are in agreement. 
For the retrieval state there is, typically, a small gap in the 
equilibrium distribution around h=0. For small $\alpha$ the gap is
very narrow.  Furthermore, in the simulations one sees that this gap
shows up very quickly. For the non-retrieval state the gap is
typically much bigger. Again in the simulations one quickly sees the gap
but it is extremely difficult numerically to find points touching the
zero  axis because of finite size effects.  

Figure 3 shows the gap width at equilibrium, $\Delta h$, for the
non-retrieval state as a 
function of $Q$ with $b=0.5$. It scales as $\Delta h \sim 1/(Q-1)$ and,
hence, decreases to zero for $Q \to \infty$. This constant behaviour of 
$(Q-1) \Delta h$ attains already for values of $Q \geq 20$ and is also
seen for the retrieval state. 
These results are insensitive to the  structure of the symmetric
architecture. 

In figure 4 the gap boundaries in $h$ as a function of $\alpha$ are
compared for retrieval and non-retrieval states in
the symmetric diluted $Q=3, b=0.2$ model. We remark that in this case  
the spin-glass states do not exist for $\alpha \leq 0.04$
\cite{BCS} so
that there is no gap for these $\alpha$-values.
For $\alpha$ large enough ($\alpha > 0.465$ for retrieval states and   
$\alpha > 0.252$ for spin-glass states) 
 there exists one gap only since the effective gain function
becomes Ising-like \cite{BCS}. More gaps  with smaller
widths are 
formed when increasing $Q$ for both the fully connected and diluted models.
For $Q \to \infty$ the gaps disappear. 
 
Figure~5 compares the gaps for the spin-glass states in the fully
connected  and symmetric diluted 
$Q=3$ models with $b=0.5$. For $\alpha \leq 0.25 $  there exist no
spin-glass states in the diluted model \cite{BCS} and for $\alpha \leq
0.004 $ there are none in the fully connected model \cite{BRS}. When
both do exist the gap widths are almost equal. So the 
dilution has some influence on the existence of the gap but, again, not
on its width. 
 
Finally, fig.~6 presents the local field distribution for the symmetric
diluted $Q=3, b=0.5$ model for a retrieval state ($\alpha =0.6, m_0=0.7$)
just below the critical capacity. Only the distribution with
pattern values $+1$ is shown. It is asymmetric and two gaps are found at
equilibrium. For pattern values $0$ the distribution is symmetric and
the gap locations and widths are the same (see eq.~(\ref{eq:distri}))
but their height is different.

In conclusion, we have studied the time evolution of the local field in
symmetric $Q$-Ising neural networks both in the retrieval and spin-glass
regime.  We have found a gap structure in the local field distribution
depending on the specific architecture and on the value of $Q$. The
results agree with the numerical simulations we have performed.

\section*{Acknowledgments}
\pin
This work has been supported in part by the Fund of Scientific Research,
Flanders-Belgium and the Korea Science and Engineering
Foundation through the SRC program. The authors are indebted to
A.~Coolen, G.~Jongen and V.~Zagrebnov for constructive discussions.

%\newpage

%\end{document}

\newpage
\pagestyle{empty}
\begin{figure}[t]
\epsfig{file=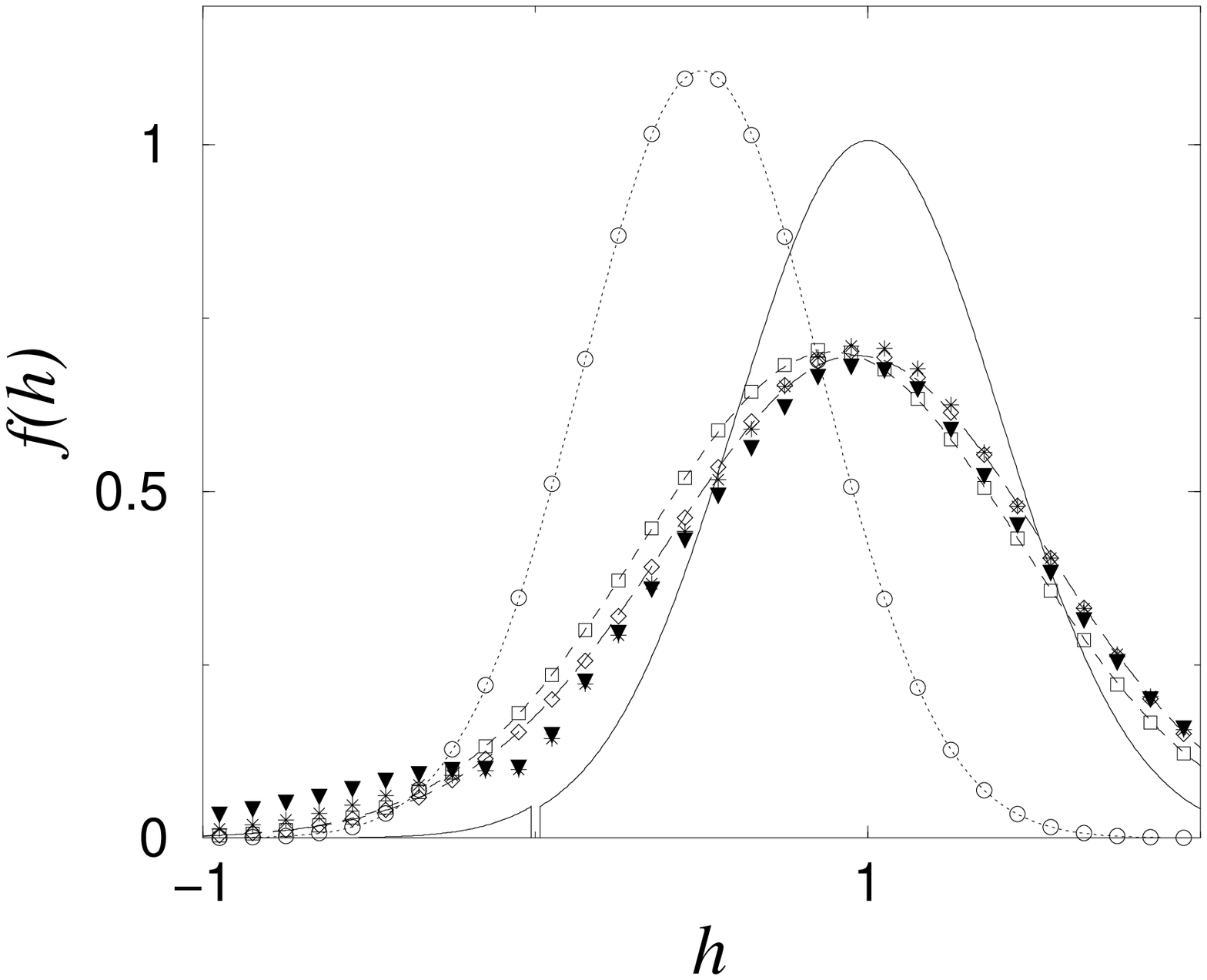, height=7.3cm}
\vspace*{-0.4cm}
\caption{\small A comparison of theoretical results and numerical
simulations  with $N=6000$ for the
local field distribution $f(h)$ of a retrieval state in the $Q=2$ system 
with network parameters $\alpha= 0.13, m_0=0.5$.  
Theoretical (simulation) results for time step $t=0,1,2$
are indicated by a dotted curve (circles), a 
short-dashed curve (squares) and a long-dashed curve (diamonds).
 Simulations for $t=10,20$ (stars, triangles) are shown and
the full curve presents the equilibrium distribution.}
\end{figure}
%\vskip 1em
\begin{figure}[b]
\epsfig{file=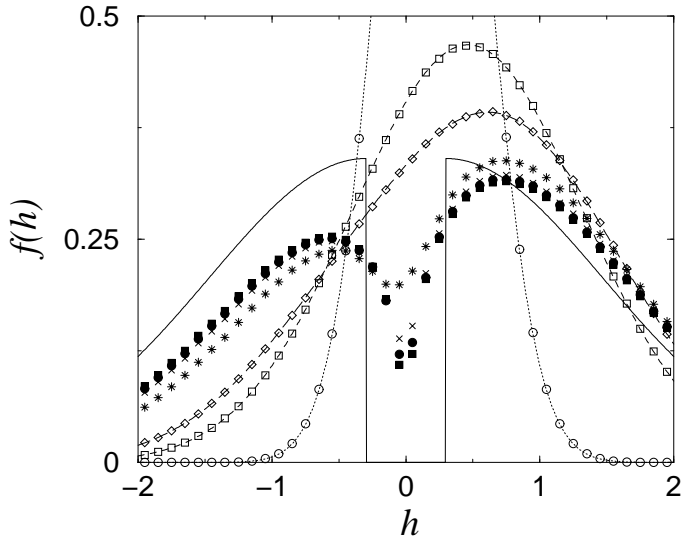, height=7.3cm}
\vspace*{-0.4cm}
\caption{\small As in Fig.~1, for a $Q=2$ non-retrieval spin-glass state with the
network  parameters $\alpha= 0.14, m_0=0.2$. Further simulations for
$t=10$ (stars), $t=30$ (crosses), $t=50$ (filled circles) and
$t=100$ (filled squares) are shown.}
\end{figure}
\begin{figure}[t]
\epsfig{file=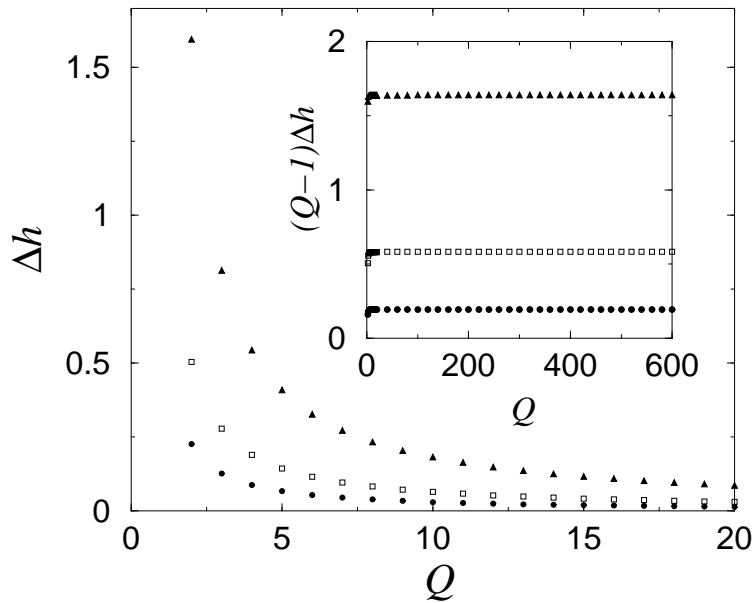, height=8cm}
\vspace*{-0.4cm}
\caption{\small The gap width $\Delta h$ for non-retrieval states as a
function  of $Q$ for the gain parameter
$b=0.5$ for $\alpha=1$ (triangles), $\alpha=0.1$ (squares) and
$\alpha=0.01$ (filled circles). The inset details the corresponding scaling
properties.}
\end{figure}
%\vskip 6em
\begin{figure}[b]
\epsfig{file=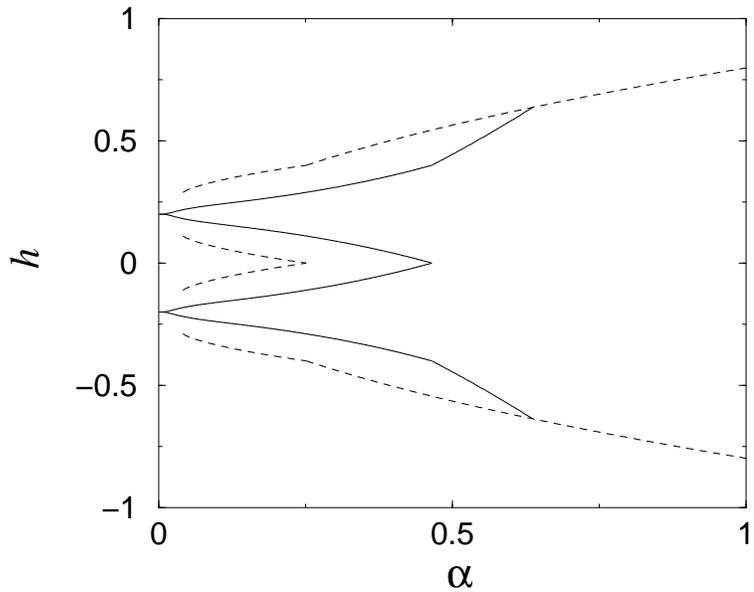, height=8cm}
\vspace*{-0.4cm}
\caption{\small The gap boundaries in $h$ as a function of $\alpha$ for 
retrieval (full
curve) and non-retrieval (dashed curve) states  for the $Q=3$ symmetric 
diluted systems with gain parameter $b=0.2$.}           
\end{figure}
\begin{figure}[t]
\epsfig{file=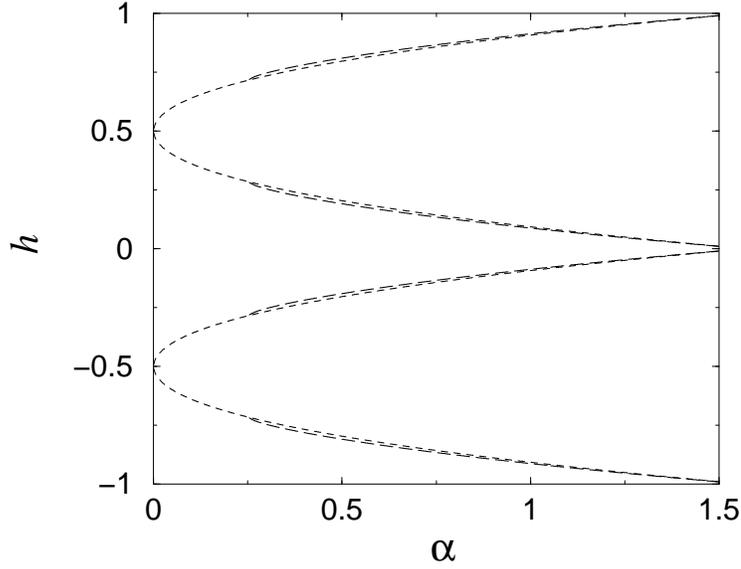, height=7.7cm}
\vspace*{-0.4cm}
\caption{\small The gap boundaries in $h$ as a function of $\alpha$ for 
spin-glass states in the fully connected  (short-dashed curve)
and symmetric diluted (long-dashed curve) $Q=3$ 
system with gain parameter $b=0.5$.}           
\end{figure}
%\vskip 6em
\begin{figure}[b]
\epsfig{file=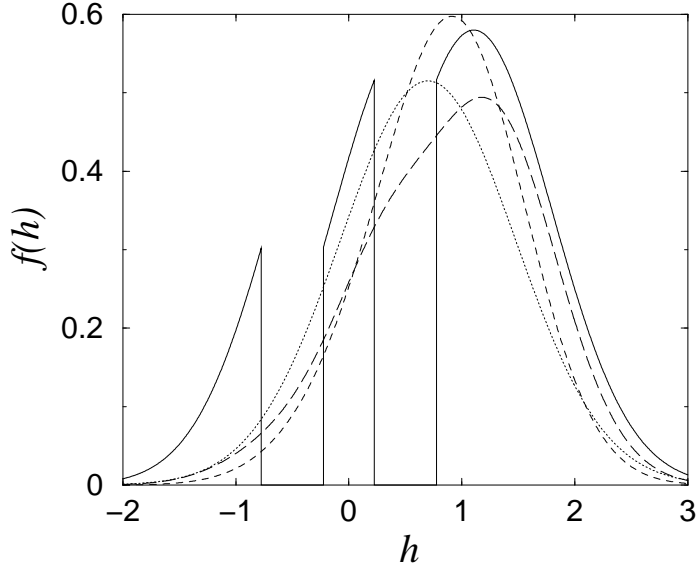, height=7.7cm}
\vspace*{-0.4cm}
\caption{\small The local field distribution $f(h)$ of a retrieval state
for  pattern values $+1$ in the symmetric diluted $Q=3$ system 
with network parameters $\alpha= 0.6, b=0.5, m_0=0.7$.  
Results for $t=0,1,2,\infty$  are indicated by a dotted curve, a 
short-dashed curve, a long-dashed curve and a full curve respectively.} 
\end{figure}

\end{document}